# Dynamic microscopic structures and dielectric response in the cubic-to-tetragonal phase transition for BaTiO$_3$ studied by first-principles molecular dynamics simulation

L. Xie,[1,2] Y. L. Li,[1,2] R. Yu,[1,2] and J. Zhu[1,2]

[1]*Beijing National Center for Electron Microscopy, Department of Materials Science and Engineering, Tsinghua University, Beijing 100084, China*

[2]*State Key Laboratory of New Ceramics and Fine Processing, Department of Materials Science and Engineering, Tsinghua University, Beijing 100084, China*

The dynamic structures of the cubic and tetragonal phase in BaTiO$_3$ and its dielectric response above the cubic-to-tetragonal phase transition temperature ($T_\text{p}$) are studied by first-principles molecular dynamics (MD) simulation. It's shown that the phase transition is due to the condensation of one of the transverse correlations. Calculation of the phonon properties for both the cubic and tetragonal phase shows a saturation of the soft mode frequency near 60 cm$^{-1}$ near $T_\text{p}$ and advocates its order-disorder nature. Our first-principles calculation leads directly to a two modes feature of the dielectric function above $T_\text{p}$ [Phys. Rev. B **28**, 6097 (1983)], which well explains the long time controversies between experiments and theories.

Materials with ferroelectric (FE) properties have long been intensively researched for their wide applications such as transducers, actuators, capacitors and memories.[1, 2] Among the mostly investigated materials, $BaTiO_3$ is of particularly interest for its typical first-order phase transitions from cubic (C) to tetragonal (T). A classical explanation of the microscopic mechanism of the phase transitions is the displacive model,[3, 4] in which all the Ti atoms stay at the center of oxygen octahedron in C phase, but will displace collectively along the [001], [011] and [111] polarization directions for the T, O and R phases, respectively. However, this model contradicts the x-ray absorption fine structure (XAFS) experiments[5] which show that the local structural environment remains approximately rhombohedrally distorted in all phases. The presence of diffuse x-ray scatterings [6, 7] in all except the R phase also suggests a spontaneous symmetry breaking and short-range ordering of local structures, which are inconsistent with the displacive model.

To explain these experiment findings that the displacive model fails to predict, Comes *et al.*[7] and Chaves *et al.*[8] proposed an eight-site order-disorder (OD) model, in which all the Ti atoms are assumed to be statically displaced along eight-possible <111> directions for all phases. In C phase, the displacements are correlated in the <100> directions and there's no macroscopic polarization on average. As temperature decreases, displacement ordering along <100> chains are sequentially established in the polarization directions and leads to the OD phase transitions. Nevertheless, the Curie-Weiss (C-W) constant of BaTiO$_3$ is 1.6×10$^5$ K, which is about two orders of magnitude large than the typical value 1×10$^3$ K for an OD type transition. The OD model also fails to predict the existence of overdamped low frequency phonon modes at the C-to-T transition observed by neutron scattering,[9, 10] IR reflectivity[11] and hyper-Raman scattering[12].

Using first-principle calculations within the local density approximation (LDA), it was predicted that the potential surface for the Ti atoms are deeper for rhombohedral <111> displacements than tetragonal <001> displacements, favoring the OD model.[13] The potential wells for the equal <111> off-center sites are relatively shallow and comparable to the transition temperature, which leads to the dynamic hopping of Ti ions between the <111> wells. Besides the ground state studies, the local structures and dynamics have also been investigated in large supercell by molecular dynamics (MD)[14] and Monte Carlo (MC)[15] approach, which successfully reproduce the correct transition sequence and reveal the presence of short-range chain-like local polar distortions even in the paraelectric (PE) phase far above $T_p$.

It's now generally accepted that BaTiO$_3$ is characteristic of both OD type and displacive type nature. However, several fundamental features of the exact roles and microscopic natures of displacive and OD components regarding the complex dielectric response during phase transition are still in question. For instance, experiments by infrared (IR) reflectivity[11] reported a saturation of the soft mode at 60 cm$^{-1}$ close to the Curie temperature $T_C$, while hyper-Raman scattering[12] resulted in a continuous decrease of the soft mode frequency to almost zero from above $T_C$. Attempts to reconcile the contradictory result lead to a two damped harmonic oscillators' model[16] and was later confirmed by first-principles MD simulation[17]. In reference 17, the two modes are attributed to the soft mode and the short-range correlation of the off-center displacements along the <100> directions. Nevertheless, there is no direct structure evidence that the less temperature-dependent peak is indeed associated with the short-range chain-like structure. In addition, the relation of the two modes behaviors to the OD type and displacive type transitions remains far from complete understanding. Hence, for a complete understanding of the dielectric response near the phase transition of BaTiO$_3$ with respect to its structure, i.e. the OD of the off-center displacements and the overdamped soft mode, accurate atomistic simulations and structure analysis are inevitably required.

In this paper, we'll attempt to (1) reveal the dynamic microscopic structures of the C and the T phase in BaTiO$_3$; (2) correlate our calculated structure results with the material's dielectric properties. Our first-principles MD simulations clearly demonstrate the evolution of the correlations of the off-center ions and the lattice vibration in the temperature range covering both the C and the T phase. The phonon dispersion relation is also calculated and a saturation of the soft mode frequency at ~60 cm$^{-1}$, which suggest an OD type mechanism for the C-to-T phase transition. The two modes feature of the dielectric function[16] is reinvestigated using the first-principles calculation results and the Girshberg-Yacoby model[18] and it's shown that the two modes could also be explained by the strong interaction between the off-center displacements and the soft mode.

Calculations are performed in the framework of the first-principles based effective Hamiltonian scheme of Ref. 15 and 19, within which the local modes $\mathbf{u}_i$ (which are proportional to the local atom displacements and associated with the optical phonon modes; see details in Ref. 19) and strain variables are solved by Newton's equation. Here, we use a system size of $L_x \times L_y \times L_z$=32×32×32 with periodic boundary conditions to undertake the MD simulations by software FERAM in the canonical ensemble using the Nosé-Poincaré thermostat. To correct the underestimation of lattice constant and phase transition temperatures inherited in the first-principles density functional theory calculations within LDA, a negative pressure $p$=-5.0 GPa is introduced in the simulation.[15] The first-principles MD simulation yielded a sequence of C-to-T and T-to-R phase transitions with transition temperatures of 320 K and 245 K. In our simulation, the dynamic local modes' configurations at 325 K and 270K are calculated with time step set to $\Delta t$=2 fs and the system thermalized to equilibrium for 100,000 time steps, after which 5,016,384 additional time steps are conducted to obtain the correlations of local modes and phonon dispersion relation.

The dynamic spatial correlations of local mode components are calculated following the general approach in describing the displacement order-disorder[20]

$$S_{\alpha\beta}(\mathbf{R},t) = \frac{<u_\alpha(\mathbf{r},t)u_\beta(\mathbf{r}+\mathbf{R},t)>}{\sqrt{<u_\alpha^2(\mathbf{r},t)><u_\beta^2(\mathbf{r}+\mathbf{R},t)>}} \quad (1)$$

where "<>" denotes spatial averages over $\mathbf{r}$, while $u_\alpha(\mathbf{r},t)$ and $u_\beta(\mathbf{R},t)$ are the local mode components along $\alpha$ and $\beta$ directions at position $\mathbf{r}$ and $\mathbf{r}+\mathbf{R}$ at time $t$. Eq. (1) describes how the local modes' OD and their vibrations are spatially correlated in space instantaneously. The separation $\mathbf{R}$ of two local modes could be specifically chosen to be three <100> vectors connecting nearest neighbors (the crystallographic directions are indexed in pseudocubic notations for the T phase). For each nearest neighbor (NN), we could further define one longitudinal correlation, i.e. $\mathbf{R}$=[100], $\alpha$=$\beta$=[100], and two transverse correlations, i.e. $\mathbf{R}$=[100], $\alpha$=$\beta$=[010] and $\alpha$=$\beta$=[001] of local modes, as is schematically shown in figure 1. The total number of the correlations of three NNs is 3×3=9 in all; but if the crystal symmetries are taken into account, the number of the correlations can be largely reduced. Apart from the NN's correlations, the short-range and long-range order of local modes are also determined from Eq. (1) by calculating the correlations as a function of distance.

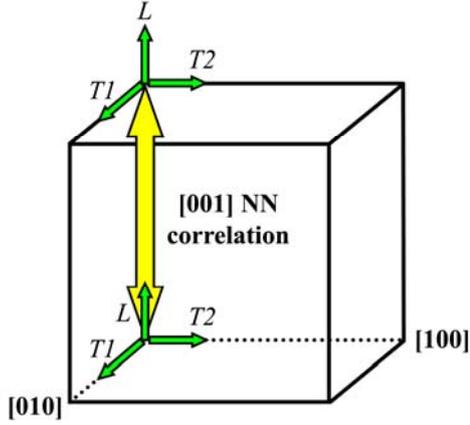

FIG. 1. Schematic view of the definition of the correlations of the local mode components. For the [001] NN, there are one correlation of the longitudinal component (L) and two correlations of the transverse components (T1 and T2).

For the C phase, its average structure is of *Pm-3m* symmetry and thus the three <001> nearest neighbors are equivalent. Figure 2(a) shows the dynamical longitudinal and transverse correlations of the [001] NN. It's clear that the [001] components of two nearest local modes are strongly correlated with an average value 0.706, while the [100] and [010] components are weakly correlated with an average value 0.187. Along the [001] direction, the [001] components' correlation $S_{[001][001]}(n[001],t)$ as a function of $n$-th [001] NN is also calculated and displayed in figure 2(b). The longitudinal correlation of the [001] components decays monotonically to zero at long range and this result suggests that along any of the <001> atomic chains, the <001> component of off-center Ti ions' displacements would preferentially form short-range positional ordering, while the perpendicular <100> and <010> components are transversely disordered along the chain's direction. This dynamic structure model is in consistent with previously reported results.[7, 14] Fitting the calculated short-range ordered curve [Fig. 2(b)] with an exponential function $\exp(-R/\xi)$, we obtain a theoretical correlation length $\xi$~3.1 unit cells, which is a little smaller than the experiment result value 5.6~5.8 unit cells.[8]

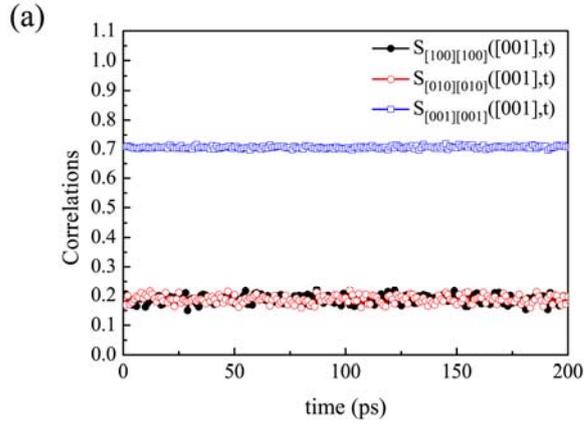

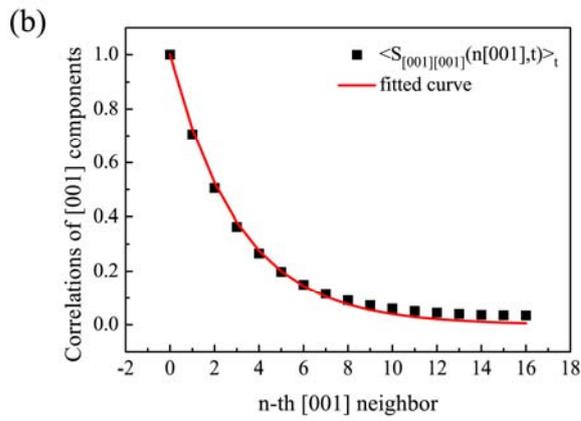

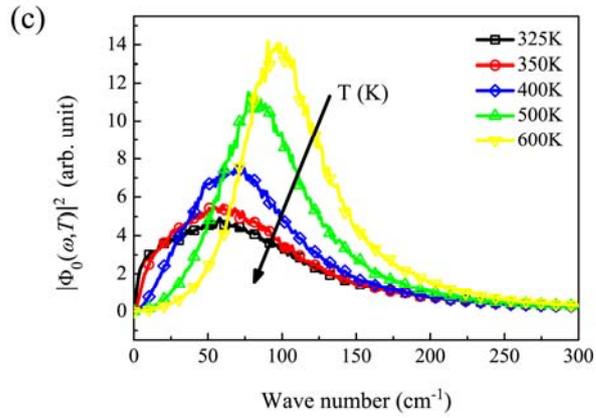

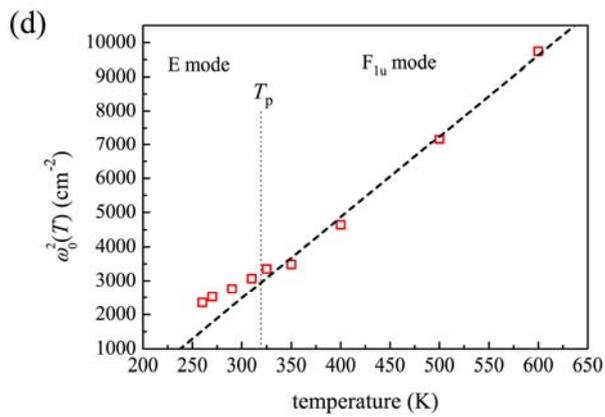

FIG. 2. Dynamic correlations of (a) the [001] NN and (b) the *n*-th [001] NN of the C phase (*T*=325 K). The solid line in (b) is the fitted result. Panels (c) and (d) display the phonon dispersion relation at Γ point and $\omega_0^2(T)$ of the soft mode in the C and T phase as a function of temperature. The dashed line in (d) is the fitted Cochran linearity relation [Eq. (4)].

Notwithstanding, it must be pointed out that the ordering may either come from the hopping between off-center sites or the lattice vibrations as the former can also be interpreted as "vibrations" of large amplitude in a grossly anharmonic potential well. For a further insight into the two different mechanisms, the phonon dispersion relation is calculated by Fourier transforming the velocity's autocorrelation function (ACF)[21]

$$\phi_{\mathbf{k}}(t,T) = \frac{<\dot{\mathbf{u}}_{\mathbf{k}}(t,T)\dot{\mathbf{u}}_{\mathbf{k}}(0,T)>}{<\dot{\mathbf{u}}_{\mathbf{k}}(0,T)\dot{\mathbf{u}}_{\mathbf{k}}(0,T)>} \qquad (2)$$

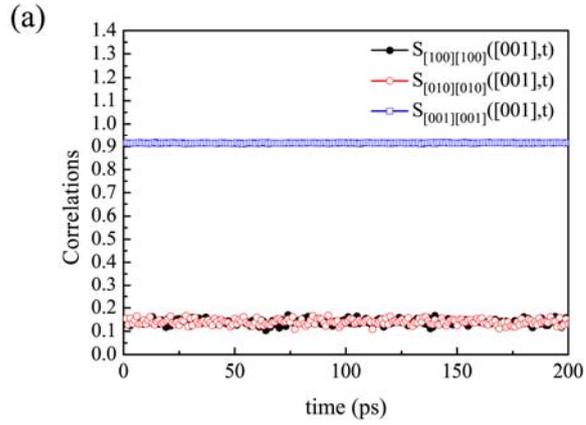

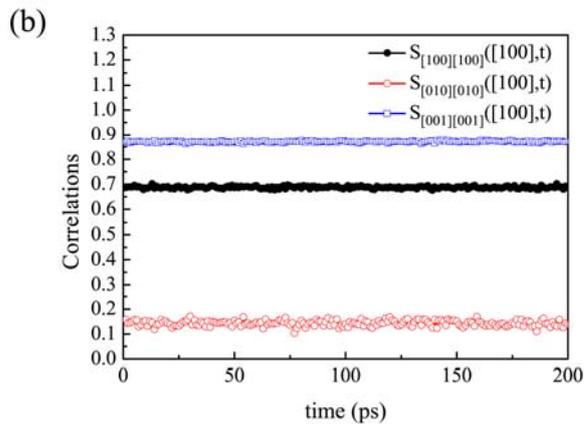

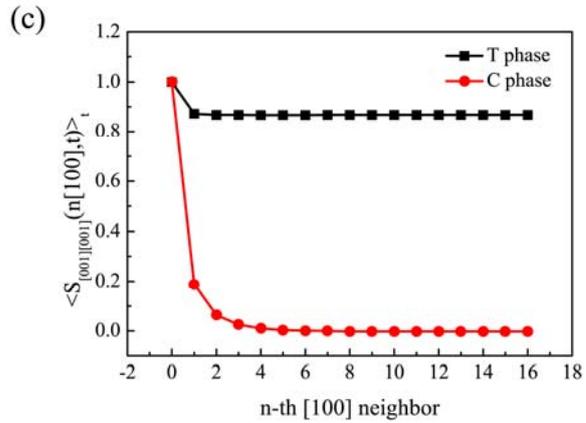

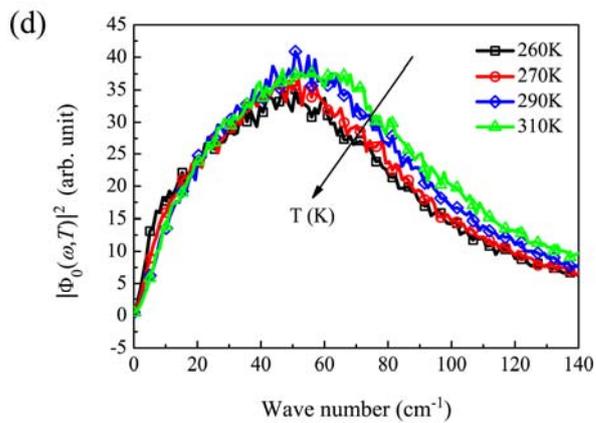

FIG. 3. Dynamic correlations of (a) the [001] NN, (b) the [100] NN of the T phase ($T$=270 K), and (c) the $n$-th [100] NN. Panel (d) displays the phonon dispersion relation of the soft E mode for the T phase.

where $\dot{\mathbf{u}}_{\mathbf{k}}(t,T) = \sum_i \dot{\mathbf{u}}_i(t,T)\exp(-i\mathbf{k}\cdot\mathbf{R}_i)$ is the Fourier transform of the local mode's velocity. In figure 2(c), we present the phonon dispersion relation at $\Gamma$ point ($\mathbf{k}$=0) for $T$=600 K, 500 K, 400 K, 350 K and 325 K. The main peak in Fig. 2(c) is strongly temperature-dependent and would correspond to the triply-degenerate $F_{1u}$ soft mode. However, no evidence of the excitation due to the short-range longitudinal correlation can be found in our calculation. From the shape of the frequency spread, the eigenfrequency $\omega_0$ of the soft mode is determined by fitting the curves with a Lorentzian function

$$\Phi_0(\omega,T) = \frac{I}{1+\left[(\omega-\omega_0)/\gamma\right]^2} \quad (3)$$

where $I$ is the peak magnitude, and $\gamma$ is the half-width at half-maximum which is dependent on the damping constant of the oscillator. The fitted results are displayed in figure 2(d). Far above $T_p$, the soft mode frequency satisfies the Cochran linearity relation[3,4] quite well

$$\omega_0^2 = A(T-T_0) \quad (4)$$

where the constant $A$ and the soft mode instability point $T_0$ is determined to be 23.8 cm$^{-2}$/K and 195.6 K. Close to $T_p$, the soft mode will become heavily damped and its frequency saturates at about 60 cm$^{-1}$, which suggests that the C-to-T phase transition is majorly governed by the OD type mechanism rather than the displacive type.

It's interesting to compare the dynamic structure and behavior of the T phase $BaTiO_3$ with respect to its C phase's counterparts. The space group of the T phase structure is *P4mm* and its macroscopic polarization is along [001] direction. According to the symmetry, the correlations of [100] and [010] NNs are equivalent, but should be distinguished from the [001] NN. As is shown in Fig. 3(a), the dynamical transverse correlation of [001] nearest neighbor is slightly decreased to 0.14, while the longitudinal correlations are enhanced by the molecular field from 0.706 to 0.915. Hence, along the [001] chains, the [100] and [010] components are still disordered in T phase as in C phase. For the [100] nearest neighbor, the [100] longitudinal correlation and the [010] transverse correlation also approximate its C phase counterparts [Fig. 3(b)]. The major difference, however, is the dramatic increase of the [001] components' transverse correlation as an immediate result of the dynamic ordering of the [001] components. As is shown in figure 3(c), the correlation of the C phase decreases monotonically to zero while that of the T phase remains almost unchanged up to long range, which thus leads to the tetragonal distortions and macroscopic [001] polarization. It may also be expected that similar mechanism holds true for the T-to-O and O-to-R phase transitions as a result of the sequential condensation of the local mode components.

Similarly, the phonon dispersion relations of the T phase are calculated according to Eq. (2). In figure 3(d), we present the $\Phi_0(\omega,T)$ at $T$=310, 290, 270 K and 260 K. The strong peak at 50-60 cm$^{-1}$ corresponds to the doubly-degenerate E mode and is correlated to the $F_{1u}$ mode in the C phase. The $\omega_0^2(T)$ of the E mode are also shown in Fig. 2(d) and it's found that the soft mode varies almost continuously from above $T_p$. Our calculation result is in excellent agreement with the IR reflectivity measurements[13], which also showed that the frequency of the soft phonon mode decreases quite smoothly across the phase transition. These theoretical and experimental consistencies confirm the OD type nature for the C-to-T transition and our calculated soft mode instability point value, which is commonly believed to be only several Kelvin below $T_C$. As the soft E mode originates from the soft $F_{1u}$ mode and persists throughout the T phase, it is reasonable to expect that the instability point of the soft E mode is quite close to that of the $F_{1u}$ mode. Hence, the instability point $T_0$ should be at least no larger than the T-to-O transition temperature. Otherwise, the E mode becomes unstable at $T_0$ and a pure displacive type phase transition would occur before the T-to-O phase transition. Therefore, our estimated value is quite reasonable since the soft mode can never reach its instability point in T phase.

However, in terms of the material's dielectric properties, how to explain the large C-W constant for an OD type transition? Moreover, although our calculated soft mode frequency shows complete agreement with the IR reflectivity, the result is however in contradict with that obtained by hyper-Raman scattering ($\omega_0 < 20$ cm$^{-1}$)[14]. The large discrepancy used to be attributed to the coexistence of two modes rather than one single mode[16, 17]. Nevertheless, as we will show below, the large C-W constant and the two modes could also be explained by the Girshberg-Yacoby model[18] with our first-principles MD simulation self-consistently. Girshberg and Yacoby first came to realize the importance of the interaction between off-center displacements and the soft mode and they described the long wavelength dielectric constant $\varepsilon$ of the off-center pseudo-spin and phonon system as a function of frequency $\omega$ and temperature $T$ as

$$\varepsilon(\omega,T) = \frac{e^{*2}}{M^* V_C \varepsilon_0}$$
$$\times \frac{(i\omega+\upsilon) + 2(\omega_0^2 - \omega^2 + 2i\omega\Gamma)(\upsilon M^* b^2/k_b T) + 4\upsilon b\sqrt{M^* \alpha_0/2k_b}/T}{(i\omega+\upsilon)(\omega_0^2 - \omega^2 + 2i\omega\Gamma) - \upsilon\alpha_0/T} \quad (4)$$

where $e^*$ and $M^*$ are the effective charge and mass of the soft mode, $V_C$ is the cell volume, $\alpha_0$ is the spin-phonon coupling (SPC) constant, $b$ is the local displacement, $\upsilon$ is the lattice relaxation rate related to the hopping mechanism, and $\Gamma$ is the damping constant of the phonon, respectively. The first, second and last term in the numerator each represents the contribution of phonon, pseudo-spin and spin-phonon coupling excitations to the dielectric function. In the absence of SPC, Eq. (4) is reduced to the simple case of two normal excitations: a lattice relaxation with rate $\omega=iv$ and a damped harmonic oscillator with frequency $\omega_0(T)$. The SPC constant $\alpha_0$ could be self-consistently determined to be $8.64\times10^5$ cm$^{-2}$ K by solving the equation[18]

$$T_C = \frac{T_0}{2} + \frac{\sqrt{T_0^2 + 4\alpha_0/A}}{2} \quad (5)$$

where $T_C$=312 K is determined from our simulation. This value is comparable to that of KNbO$_3$[18] ($6.3\times10^6$ cm$^{-2}$K) but is much larger than the value obtained by Pirc and Blinc[22] ($4.096\times10^3$ cm$^{-2}$K) for BaTiO$_3$. Such large discrepancy is quite straightforward. As can be seen from Eq. (5), the smaller the difference between $T_0$ and $T_C$, the smaller the SPC constant $\alpha_0$ is. As we have discussed above, the instability point should not be too close to $T_C$, otherwise the tetragonal structure becomes unstable immediately after the C-to-T transition. From our MD calculation, all the parameters used in Eq. (4) could be obtained *ab initio* except for the hopping rate $v$ and the damping constant $\Gamma$ of the soft phonon mode. In the limit of zero frequency, the dielectric constant is irrelevant to $v$ and $\Gamma$:

$$\varepsilon(0,T) = \frac{e^{*2}}{M^*V_C\varepsilon_0} \frac{1+2\omega_0^2 M^* b^2/k_b T + 4b\sqrt{M^*\alpha_0/2k_b}/T}{\omega_0^2 - \alpha_0/T}$$
$$\equiv \frac{C}{T-T_C} \quad (6),$$

and we obtained the C-W constant $C=2.54\times10^5$ K, which is in quite good agreement with the experiment value. Using the hopping rate $\upsilon$ determined by J.-H. Ko *et al.*[23]: $\upsilon = \beta(T-T_{OD})$ with $\beta$=1.08 cm$^{-1}$K$^{-1}$ and the shifted OD transition temperature $T_{OD}$=263 K[24], and assuming that the damping constant $\Gamma > 50$ cm$^{-1}$, the theoretical dielectric functions [Eq. (5)] are calculated for $T$=350 K, 400 K and 500 K and are shown in figure 4(a). Although there is only one soft mode, it's obviously that there is a main peak at 10-20 cm$^{-1}$ as well as a broadened peak at 50-60 cm$^{-1}$ in the imaginary part of $\varepsilon(\omega,T)$. As is approaching $T_p$, the main peak will shift to lower frequency while the second peak is less temperature dependent. This two modes feature exactly reproduces the experiment findings but has a completely different origin, which formerly was attributed to the excitation of the soft mode and the short-range correlation of <100> chains[17]. As is shown in figure 4(b), the dielectric function [Eq. (4)] is decomposed into individual responses of phonon, pseudo-spin and spin-phonon coupling terms: $\varepsilon(\omega,T) = \varepsilon_{ph}(\omega,T) + \varepsilon_{spin}(\omega,T) + \varepsilon_{spin-ph}(\omega,T)$. For $\omega<\omega_0$, the large SPC constant $\alpha_0$ in $\varepsilon_{ph}(\omega,T)$ and $\varepsilon_{spin-ph}(\omega,T)$ has a dramatic effect on the dielectric response and gives rise to the observed main peak at low frequency. The main peak doesn't coincide with any damped harmonic oscillator and significantly deviates from its uncoupled case ($\alpha_0$=0). The second broad peak (50-60 cm$^{-1}$), as is indicated by the arrow in Fig. 4(b), however, comes from the real phonon mode excitation but is buried under the tail of the first peak since it's highly damped.

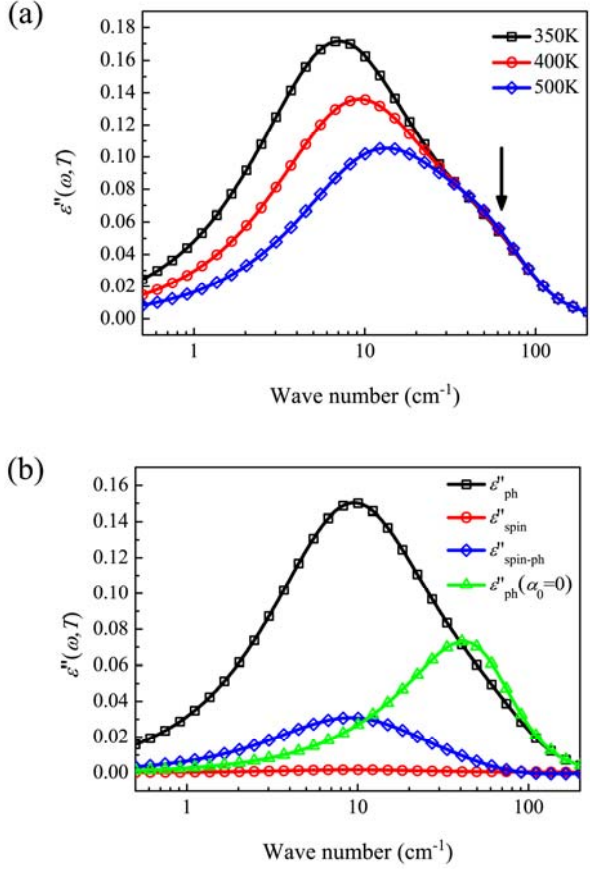

FIG. 4. (a) The imaginary part of $\varepsilon(\omega,T)$ for $T$=350 K, 400 K and 500 K calculated by Eq. (4). The arrow indicates the second mode which corresponds to the soft mode. (b) The dielectric response of phonon $\varepsilon_{ph}(\omega,T)$, pseudo-spin $\varepsilon_{spin}(\omega,T)$ and spin-phonon coupling $\varepsilon_{spin-ph}(\omega,T)$ for $T$=350 K. The line with upper triangle is the dielectric response of the uncoupled damped harmonic oscillator model.

These findings lead directly to the conclusion that the large C-W constant and the two modes feature is essentially the result of the strong interaction between off-center displacements and the soft mode. This strong coupling is formerly unexpected but is clearly shown here.

Moreover, it might also raise questions that if fitting the experiment dielectric response with harmonic oscillator model is appropriate to determine the soft mode behavior for such kind of material. Therefore, a direct observation of the overdamped phonon mode with highly accurate and sensitive techniques would be inevitably required.

As a summary, our MD studies lead directly to the dynamical microscopic pictures of the C and T phases and the transition mechanism in $BaTiO_3$. In the C phase, local mode components are dynamically short-range correlated in longitudinal directions and are less ordered in the transverse directions. In the T phase, strong long-range transverse correlations of the [001] components is established, while the [100] and [010] components remain short-range ordered in longitudinal directions and less correlated in transverse directions. Concerning with the phase transition and the dielectric properties, the phonon frequency at Γ point is calculated and it's found that the frequency of the soft mode saturates at ~ 60 cm$^{-1}$ near $T_p$, advocating the OD type transition mechanism and in consistent with the ordering of [001] local mode components. Using the Girshberg-Yacoby model with the parameters determined by our first-principles MD calculations, it's revealed that the strong coupling of off-center pseudo-spin and soft mode distinctively contributes to the low frequency dielectric response, giving rise to the large C-W constant and the two modes feature. These thorough understandings of the microscopic structures and their relation to the dielectric properties of $BaTiO_3$, which we expect, will not only benefit the optimization of devices based upon $BaTiO_3$. But also, as similar structure is ubiquitously found in many $BaTiO_3$-based type solid solutions, e.g. $Ba(Ti_{1-x}Sn_x)O_3$, $Ba(Ti_{1-x}Zr_x)O_3$ etc., whose properties could be easily tuned from normal FEs to relaxors by changing the substitution *x*, our current studies would be important in understanding the properties and the underlying mechanisms in such kind of materials as well.

ACKNOWLEDGEMENT

This work is financially supported from National Nature Science Foundation of China and State Key 973 Project of China. The authors would like to thank Dr. Nishimatsu for helpful discussions on using the MD simulation program.

[24]The hopping rate $v$ related to the critical slowing down for the OD type transition is given by $v = v_0 \dfrac{T - T_{OD}}{T_{OD}}$ and the parameters are determined to be $v_0$=370.3 cm$^{-1}$ and $T_{OD}$=343 K (30 K below the experimental C-to-T phase transition)[23]. At high temperature, the hopping rate saturates at around 63 cm$^{-1}$. In order to take into account the underestimation of the $T_p$ in our first-principles MD simulation, the function is shifted to a new temperature (30 K below the calculated $T_p$) without changing the slope $\beta = \dfrac{v_0}{T_{OD}}$, which attempts to ensure that the hopping rate is in a reasonable range in the calculation.